\documentclass[12pt,english]{article}
\usepackage{graphicx}
\usepackage[pt154]{inputenc}
\usepackage[T2A]{fontenc}

\begin{document} 

\begin{center} 
{\Large\bf Thermodynamics of economic inequality} 

\vspace{3mm}

{\large\it Vladimir Pokrovskii}\footnote{Corresponding author:  Vladimir Pokrovskii, vpok@comtv.ru} \\

\vspace{3mm}

%{Independent reasearcher, Moscow,  Russian Federation (No affiliation)}

{Moscow State University of Economics, Statistics and Informatics}  

\vspace{5mm}

\end{center}

\centerline{Abstract}

\vspace{5mm}

 An  heuristic  model of the society, as an assembly  of weakly interacting  individuals, is discussed.  The model allows to  connect macroscopic phenomena with features  of relations between individuals. Addressing to the problem of inequality, a  non-equilibrium situation is considered. The calculated   income distribution function, which  coincides with the Pareto distribution at  large incomes, can be interpreted  as a strongly deformed Gauss distribution function. The external, in relation to the system of interacting individuals, force is necessary to maintain the strong non-equilibrium in a stationary state. The model representation of the society allows us to explain  the mechanism of the emergence and maintenance of economic inequality,  the universal cause of  which is asymmetry  of elementary exchanges.

\vspace{5mm}
\vspace{5mm}

%Author e-mail  \\     vpok@comtv.ru \\

%Author affiliations   \\
%{Moscow State University of  Economics, Statistics and Informatics (MESI),\\ %Moscow, Russian Federation 119501}\\

%ORCID iDs \\
%ladimir Pokrovskii https://orcid.org/0000-0002-9358-2382 \\

%Dates

%$JEL$ - classification: C13,  C51,   D31,  D63,  D9	
\vspace{5mm}

%\centerline{February 2023}

\newpage

 \section{Introduction}

The modern  human economic activity is based on the flows of solar energy, which passes through plants, animal bodies, the work of machines and, in the end, is used by man to maintain his life. The production system was developed by the human populkation on the Earth.  The current  pattern  of economy have  been leading to significant inequality in the distribution of income and wealth [1].  The typical pattern  was uncovered by the sociologist and economist Wilfredo Pareto,  who  found ([2], pages 299-345) that the distribution density function of the individuals $p(x)$ over  income $x$ in the region  of the large values is described  by a power function
\begin{equation}
p(x) = Ax^{-(1+\alpha)}.
\end{equation}
The power law  has been verified over and over again since then for different countries and at different times; the reliable estimates of the empirical values of  index $\alpha$ in the power distribution (1), according to ([3] p.84), range from 1.2 to 3.2.
 
Since Pareto's time, many researchers have sought to understand whether the power distribution is a law of nature, or whether it can be reduced to more fundamental principles. Numerous investigations,  devoted to description and explanation the mechanism of emergence and existence of the inequality, recently are reviewed by a Brazilian researcher [3]. However,   the problem of finding the shape of the distribution function for those who do not belong to the rich class (the vast majority of people in modern societies), according to ([3], p.\,64), is remained basically open. This statement is confirmed in the paper  [4]:  'Despite decades of intense research and  data analysis, the exact physical mechanisms of wealth  distributions are not yet fully understood'.

To reconsider the problem and to try to understand  the original  of inequality we use a new approach, defining   the society to be  an open thermodynamic system.  The addressing to the a simplest  heuristic model of a market economy, as an assembly of interacting economic agents, allows us, referring to the empirical Pareto index $\alpha$, to  restore  the entire distribution density function. The model allows to   explain   the mechanism of emergence and existence  the wealth  inequality.

\section{Heuristic model of the society}

The productive activities apparently underlies the entire human life, so that, under the desire to construct an heuristic model of the society, we  are encounting, first of all,  the problems of description  of economic activity, which involves the coordinated activity of a huge number of people, who unite in large groups to manufacture needed things. Investigating the production-consumption phenomena, the total population should be taken into account, but theory, nevertheless, identifies a group of people who are directly involved into  production --- {\it economically active individuals}; we consider an assembly of  $N$  such  active  individuals, who  jointly create the entire social  product. 

 The assembly of interacting individuals,  together  with their artificial environment, ought to be  considered as an  open thermodynamic system; the existence of the system is supported by streams of matter and energy coming into the system through the bodies of   individuals and various appliances. There are two  fluxes of external energy, which, during the process of  production, feed  efforts of individuals, the total of which is denoted as $L$, and    substitutive work of production equipment $P$. These two quantities   determine production of value in money units [5, 6] 
\begin{equation} 
Y = Y_0 \, \frac{L}{L_0}\left(\frac {L_0}{L} 
\frac{P}{P_0}\right)^\alpha, \quad 0 < \alpha < 1,
\end{equation}
where the zeroes indicate the values of output, labor  and substitutive work in the base year. The time-dependent quantity $\alpha$ --- the technological index --- is an internal characteristic of the production system, and is expressed  through the technological characteristics of the production equipment\footnote{Note that the relation (2) has the form of the famous Cobb-Douglas production function, but in which, after reasoning about the mechanism of substitution of labor by capital [5], a variable $P$ is used  instead of a variable capital  $K$ that is the estimate of value of all production equipment. This led to a accurate  determination of the production of value through production factors without any corrections and arbitrary parameters [6]. Capital (production equipment) $K$ is the  means, by which the active factors of production, $L$ and $P$ can unrestrictedly  substitute each other under modification of production processes. The equipment itself is dead until 
it is not animated by streams  of human efforts and external energy. }. 

The incoming energy has been dissipated in the  production processes, and, it is assumed for simplicity,  the created  product is completely consumed by the  individuals, so that the stream of value $Y$  is coming through the  system. As in every open thermodynamic system [7], the conditions for creating of internal structure of the system appear, and, indeed,  in following sections,  we will follow a mechanism of  emergence of non-equilibrium of  the system.  

The  two production factors: the  activities of people $L$ and  substitutive work of production equipment $P$ appears to be essential in  production of value  that is an estimate of all produced goods and services. People's work is used directly, but in large-scale machine production, people's work is alienated: this means that a person must perform prescribed actions without realizing the true goals of the work, for which he receives a reward, the correspondence of which to the efforts expended is not obvious. Substitutive work is based on the preliminary activity of people for centuries in the  study of the laws of nature, the  design of technological schemes, the manufacture of production equipment and everything else. The results of this activity are a common heritage that we have received from previous generations. 

The individual contributions into the production of value are  various; in the simplest case, one can assume that the input of each agent is equal and corresponds to the mean estimate  of the  individual efforts $Y/N$. In a more complicated case, the  individual contributions could be described by the Gauss distribution.   Each participant  receives a share of the created  product --- the income of an individual $x$, which should be considered as a random variable, and, therefore, it is possible to introduce a distribution function of agents over  income $p(x)$, such that the value of $p(x) \Delta x$ represents the number  of people with income is  situated between $x$ and $x + \Delta x$. Obviously, the following relations have to be  fulfilled
\begin{equation}
\int_0^\infty   p(x) d x = N, \quad  
\int_0^\infty  x  p(x) d x = Y
\end{equation}

The market mechanism presupposes free exchange between various economic agents: the efforts of the workers  for money, money for products, products for other products, and so on. We assume that any interactions could be reduced to exchange of value and, consequently,  each individual is prescribed with  the only status variable  --- the income of an individual $x$, which  changes its value in a random way. Considering the  chaotic interactions, one can  introduce  the probability of the change of agent's status  from $x$ to $x-a$, whereas the alteration  $a$  is supposed to be distributed according to the  standard Gaussian function  
\begin{equation}
w\left(a, \langle a \rangle, \langle a^2 \rangle \right) = \left(\frac{1}{2\pi \langle a^2 \rangle} \right)^{\frac{1}{2}} \exp{\left[ -\frac{(a - {\langle a \rangle})^2}{\langle a^2 \rangle } \right] } 
\end{equation}
It is assumed  that the transition probability function could depend on the variable $x$ through the  mean value ${\langle a\rangle}$ and standard deviation ${\langle a^2\rangle}$.

This simple heuristic model takes into account the multitude  of exchanges between economic agents according to  market rules in the processes of production and distribution of value and  allows us to find the distribution function of economic agents over  the income received,  which allows  to clear up the mechanism of occurrence and maintenance of economic inequality.

\section{Derivation of the distribution function}

\subsection{Kinetic equation}

We assume that society is represented by the system  of weakly interacting economic agents, the macroscopic state of which is determined by the distribution density function of the individuals over income $p(t, x)$, which depends also  on time $t$.  To find this  function by  the implementation of the described model of the assembly of individuals, changing their states due  to the random elementary transitions from one state to another, we use a known method that allows us  to write a kinetic equation for the distribution function (see, for example, ([8], chapter~2). In this case, the change in the distribution function can be written as
\begin{equation}
\frac{\partial p(t, x)}{\partial t} = 
\int_{-\infty}^{+\infty} \left[ w(a, x+a) p(t, x+a) - w(a, x) p(t, x) \right] da \end{equation}
We assume that the change in the income $a$ is  much smaller than its current value, $a\ll x$; based on this, the first term in the integrand (5) can be replaced by the first terms of  expansion 
$$
w(a, x+a) p(t, x+a) \approx w(a, x) p(t, x) +  a \left. \frac{\partial (w p)}{\partial x}\right|_{a=0} + \, a^2 \left. \frac{\partial^2 (w p)}{\partial^2 x}\right|_{a=0}.
$$
The variables $a$ and $x$ are considered independent, so that  the kinetic equation for the distribution function (5) can be written as
\begin{equation}
\frac{\partial p(t, x)}{\partial t} = \frac{\partial }{\partial x} \left[ \left(\langle a \rangle +  \frac{\partial \langle a^2 \rangle}{\partial x} \right) p + \langle a^2 \rangle \frac{\partial p}{\partial x} \right].
\end{equation}
Under these assumptions, the equation is valid in the region of positive values of $x$. If the function $w(x, a)$ is symmetric with respect to the variable $a$, $\langle a\rangle = 0$, and the probability variance does not depend on $x$, the first term on the right side of equation (6) disappears, and the kinetic equation reduces to a one-dimensional diffusion equation. To describe more complex situations, it should be assumed that the transition function could  be asymmetric with respect to $a$, and the average value and variance  may depend on the variable~$x$. 

The square brackets on the right side of the equation (6) contain two terms: the diffusion leveling  is compensated by the flow. The equation  describes a balanced internal motion and defines a non-equilibrium distribution function, what is supported by external streams of matter and energy.   

\subsection{Steady-state distribution}

In the steady-state case, the terms of expression (6) enclosed in square brackets does not depend on $x$, and therefore we can introduce a constant $C$ and write the equation 
 \begin{equation}
\frac{d p}{d x} =\frac{1}{\langle a^2 \rangle}  
\left[C - \left(\langle a \rangle +  \frac{d \langle a^2 \rangle}{d x} \right) p \right]. 
\end{equation}
We are looking for a solution of this equation, which, for large values of $x$, should  correspond to the Pareto distribution (1). The limit value of the function $p(\infty)$ is not known, but an expression for the derivative can be set as 
\begin{equation}
\left(\frac{d p}{d x} \right)_\infty = - \hat{p} (1+\alpha) x^{-(2+\alpha)}, \quad p = p(\infty) +  \hat{p} x^{-(1+\alpha)}.
\end{equation}

The  mean value ${\langle a \rangle}$ and the variance of ${\langle a^2 \rangle}$ in  equation (7), depend on income in the region of positive values of $x$. It can be expected that these quantities  are increasing functions of the values of $x$ (otherwise, exchanges of agents with high income may be inefficient), which are conveniently represented with power function. To satisfy  the asymptotes (8), we choose the dependencies
\begin{equation}
\langle a \rangle = \langle a \rangle_0 + r x^{1+\alpha}, \quad \langle a^2 \rangle = \langle a^2 \rangle_0 + k x^{2+\alpha}. 
\end{equation}
With these values of the indices, equation (7) has  the desirable  asymptotic (8),  and value of the constant in  equation (7)  could be found
\begin{equation}
C = \hat{p} \, [1 + r + (1+k)(1+\alpha)] , \quad   p(\infty) = 0. 
\end{equation}
%\linebreak
%\vspace{3mm}
%\hrule
%{\setlength{\unitlength}{1mm}
%\vspace{1mm}
\hspace*{4mm}
\centerline{\includegraphics[scale=0.5, bb=0 0 650 350]{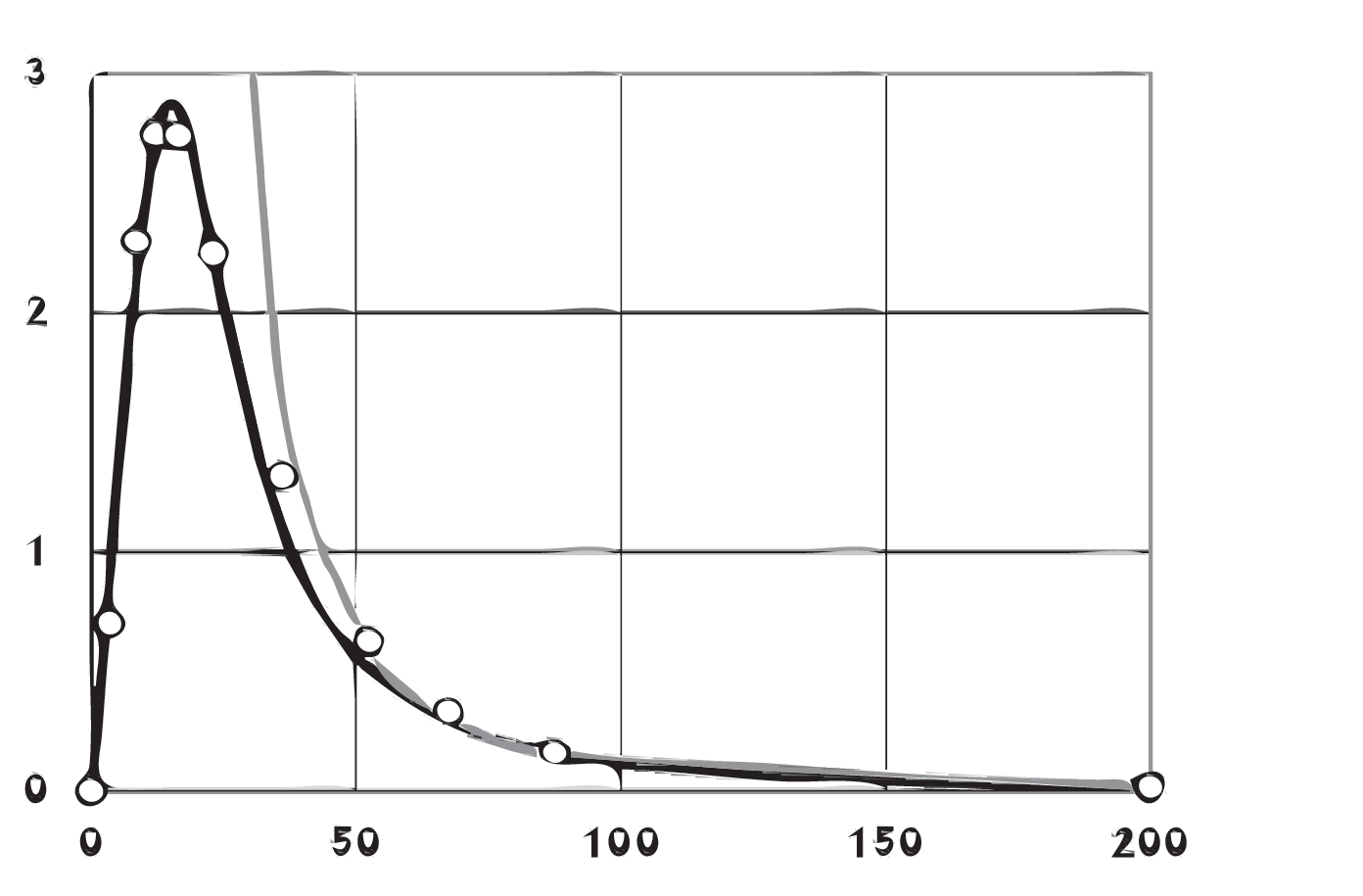}} 

\vspace{5mm}

\centerline{\bf \ Figure 1 Distribution density function}
\vspace{2mm}
{\it\small\noindent The empty circles show empirical values of the density of distribution of the Russian population by income in 2018, calculated  according to available data [9]. The slight line represents the Pareto distribution (1) at $\alpha = 1.65$. The solid line shows  the solution of equation (7) at the values of the parameters (11) and (12). The distribution density function is normalized by 100 units. Income is  shown  in thousands of rubles.}

\vspace*{3mm} 
\hrule 
\vspace*{6mm} 
%\noindent
Equation (7) with  relations (9) - (10) appears to be  an ordinary differential equation of the first order with the initial condition $p(0) = 0$, which is impose additionally.

 Without looking  for an analytical expression of the solution, we turn to numerical methods that we use for the specific case of the distribution of individuals over income for Russia in 2018, taken as an example.  In  Figure 1, the empty circles represent the values of the probability density of the distribution estimated  according to Rosstat data [9]. One can notice that the dependence represents  typical income distributions,  as could be seen  from the graphics on pages 93 and 98 of the review [6]. In  Figure 1,  a slight line shows  the Pareto power law (1) with index $\alpha = 1.65$, which reproduces empirical results  above the income $x > 50$.

Equation (7) is considered as an ordinary differential equation, which we solve numerically by a standard method, taking into account relations  (9).  The simple consideration allows us to choose the parameters: for zero   values of  income, the mean values are  close to zero, whereas the variance is not zero. 
\begin{equation}
\langle a \rangle_0 = 0; \quad \langle a^2 \rangle_0 = 0.02. 
\end{equation}
The quantities increases with increasing income according to the law (9) with
the parameters, which must obey to some conditions.  Both quantities did not exceed the total amount of income, which determines the small values of the parameters
\begin{equation}
 r =  0.0000014; \quad \quad  k = 0.0000005. 
\end{equation}

The numerical values in the ratios (11) and (12) are selected in such a way that the maxima of the theoretical and empirical dependencies coincide. The value of constant $C$ is determined by the relation  (10), while  the parameter $\hat{p}$ can take an arbitrary sufficiently large value; calculations are carried out at the  value of $C = 36.6$.

We are looking for a solution of  equations (7) - (10) with the values of the parameters (11) and (12) that are specific values for every sample.   The result of calculating the distribution function is shown in  Figure 1 in comparison 
with the empirical values and the Pareto distribution. This result convinces 
us that the entire distribution function could be restored according to the Pareto index and empirical values of the maximum of income and its  position in relation to the mean income.

\section{The anatomy of inequality}

Returning  to the consideration of an assembly of individuals, each of whom, according to our simplest assumption, produces the  $Y/N$ units of value (see Section 2),  so that distribution of individuals over  participation in production  could be represented  by delta-function. More exatly, the distribution could be represented by  the standard Gaussian distribution about  the mean  value of income $Y/N$. 
 
The distribution of individuals over  income could  be different depending on the circumstances of the allocation. In a simple case, when there are only random interactions of individuals and other chaotic uncontrolled influences, the probability function of the distribution of individuals over  income takes the form of the standard Gaussian distribution about  the mean  value of income $Y/N$
\begin{equation}
p\left(x \right) = \left(\frac{1}{2\pi \langle (\Delta x)^2 \rangle} \right)^{\frac{1}{2}} \exp{\left[ -\frac{(\Delta x)^2}{\langle (\Delta x)^2 \rangle } \right] }.  
\end{equation}
\linebreak
%\vspace{3mm}\
%\hrule

%{\setlength{\unitlength}{1mm}
%\vspace{1mm}
%\hspace*{4mm}

\centerline{\includegraphics[scale=0.28, bb=0 0 1050 550]{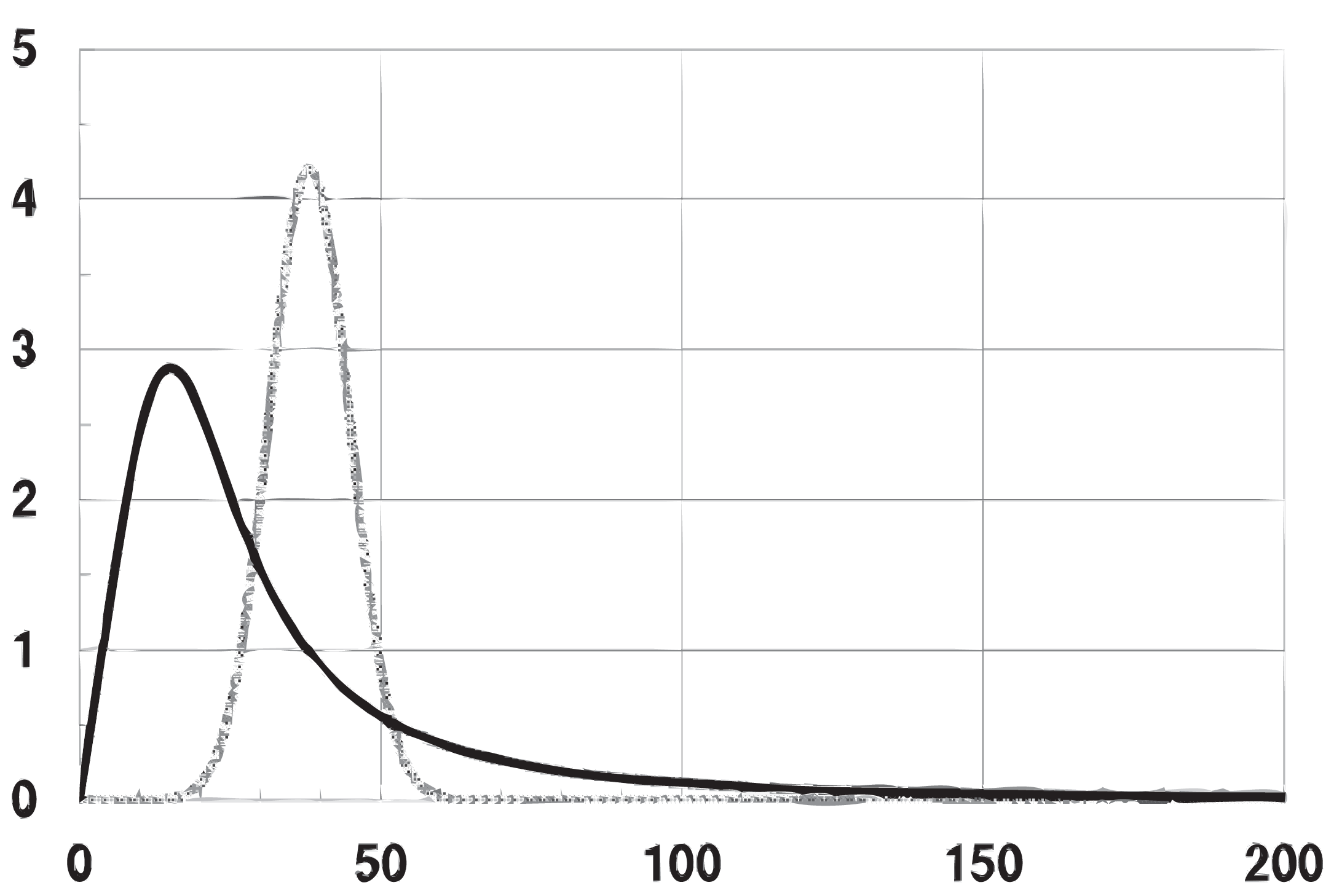}}%

\vspace{5mm}

\centerline{\bf Figure  2 \ Income distribution functions}
\vspace{2mm}
{\it\small \noindent The slight  curve represents the natural distribution of individuals over income, calculated for  the average value of contribution  by an individual $Y/N=38.1$ and variance $\langle (x - Y/N)^2 \rangle = 90$. The solid curve represents the real distribution of the Russian population by income in 2018, reproduced from Figure 1. The distribution functions are normalized by 100 units. Income is presented in thousands of rubles.}
\vspace*{3mm} 
\hrule 
\vspace*{6mm} 
\noindent
Here it is denoted $\Delta x = x - Y/N$. The graph of the  distribution (13) is shown in Figure 2 by the slight line. 

In reality, the situation appears to be more complicated. The  share of the created product, which receives by an individual,  does not correspond,
generally speaking, to the contribution of the individual to production of value.  The graph of the real distribution function of individuals by income, reproduced in  Figure 2 with the solid line,  shows that  the real distribution function of individuals over income differs significantly from the Gaussian distribution and has the form (1) in the region of big incomes.  The real distribution looks like a strongly deformed Gaussian distribution.  Both one and the other distributions shown in Figure 2 are stationary non-equilibrium distributions.

 Turning to the situation in more detail, we note that, according to the assumption,  each individual, regardless of his real income, produces $Y/N = 38.1$ units of value. It can be easily estimated that about 70\% of individuals 
have a real income less than the average value  $Y/N =38.1$; in total they receive about 35\% of the total product. The remaining part of individuals, respectively, receives an income exceeding the mean  value, which means that there is a mechanism that transfer some amount  of the value from individuals with lower income to individuals with higher income. Such a flow of value cannot arise naturally, without some influences; there is an embodied mechanism, which  supports the  trends towards the non-equivalence of elementary exchanges: in  each act, the agent who already receives the higher income has an advantage. \  Formally,\ the non-equivalence \ of the elementary exchanges is related to the asymmetry of the  transition function (4) between states with a non-zero mean transition value.

The description of inequality in income  distribution were discussed by many autors [6].  Yakovenko and his collaborates, for example,   noticed  that the empirical distribution functions can be represented as a composition of two functions [10]: for small values of the variable, the distribution is described by the Boltzmann-Gibbs exponential function, while for the large values of the variable, the distribution corresponds to the Pareto power law. This gives them a reason to present the totality of individuals as quasi-independent communities of the poor, the number of which, according to Ludwig and Yakovenko [10],  surpasses 90\% in the USA, and the rich (about 4\% in the USA). Not to mention the fact that the two-component representation of distribution causes an feeling  of professional dissatisfaction, the quasi-equilibrium two-component scheme of Ludwig and Yakovenko omits the essence of the phenomenon: the mechanism of interchange between the poor and the rich, in the process of which the wealth flows from the poor to the rich, and, one can think that  the representation of distribution by a single function gives a more adequate picture of the phenomenon; in any case, it is useful to look at the entire function.

\section{Conclusion}

The representation of the society  as an  assembly of weakly interacting individuals appears to be  a conventional beginning of  all models, aspiring to connect macroscopic properties  of the society with features  of relations between individuals [3, 4]. A peculiarity of the presented  approagh is  considering  the assembly of individuals to be an open thermodynamic system that cannot exist without external streams of energy through bodies of the individuals and production equipment.  The law of production (2), determining   relationship between the production of value and production factors is objective law, this relation  cannot be changed according to the desires of people,  and there is nothing in tht scheme of production  that hints on the inequality. The inequality is connected with peculiarities of behavior of economic agents; the distribution of income can be explained by asymmetry of elementary exchanges that is external in relation to production processes.  In contrast to the laws of production, the distribution of products is  determined by rules that are established by people based on ideas about the production of value, taking into account production factors. The formulated rules are then reformulated into laws, approved by Parliament, written into the constitution and implemented: the rules for the distribution of social product are legitimized, and the life of society (social relations) is organized in such a way as to guarantee the implementation of the written rules. The difference in social systems is significantly related to the rules of distribution of the public product, which need to be revised from time to time.

 In the simple case, when  streams of matter and energy into and out the society  are ignored,  the  model system is   considered  to be  in the equilibrium state, which allows to define equilibrium distribution functions. However, one ought to be conscious: there is the exchange of matter and energy with the environment, and the society  ought to be considered as an open thermodynamic system that cannot exist without external streams. This paper demonstrate how the usage of a thermodynamic model helps to uncover mechanism of the emergence and maintenance of economic inequality. The knowledge of the mechanism allows us to determine  recipes for mitigating the inequality, but this is a different  problem that  could  be considered on the basis of the results of this investigation. 

\newpage

\section*{References}

\begin{enumerate}

{\small 

\item Piketty T. (2014)   {\it Capital in the Twenty-First Century.}  Harvard University Press, Cambridge 

\item Pareto V. (1897).   {\it Cours d'economique politique. The first edition}.  Macmillan, London, Also: Pareto V. {\it Cours d'Economie Politique: Nouvelle edition \linebreak 
par G.-H. Bousquet et G. Busino}, Librairie Droz, Geneva, 1964. 

\item Ribeiro M.B.  (2020). {\it Income distribution dynamics of economic systems: An econo-physical approach}. Cambridge,  UK:  Cambridge University Press.  
\item Aydiner, E., Cherstvy, A.G. and Metzler R. (2019) Money distribution in agent-based models with position-exchange dynamics: the Pareto paradigm revisited?    Eur. Phys. J. B (2019) 92: 104 https://doi.org/10.1140/epjb/e2019-90674-0 

\item Pokrovskii, V.N. (2018)  {\it Econodynamics: The Theory of Social Production, 3nd Ed.}  (Dordrecht-Heidelberg-London-New York: Springer) 

\item  Pokrovskii, V.N. (2021) Social resources in the theory of economic
growth.  {The complex systems}, № 3 (40), 33-44. Available at: \\ https://www.researchgate.net/publication/355163496/  (accessed: 16 September 2023).

\item Pokrovskii V.N. (2020)  Thermodynamics of Complex Systems: \  Principles and  applications.  Bristol (UK),  IOP Publishing, 2020.  

\item Lifshitz E.M. and Pitaevskii L.P. (1981) {\it Physical Kinetics}. Oxford: Pergamon. 

\item Rosstat (2023).   {\it Population by average per capita money income}. Table: urov\_31.xlsx (updated 26.12.2019), Available at: https://www.gks.ru/ (accessed: 16 February 2023).

\item Ludwig D. and Yakovenko V.M. (2021).  Physics-inspired analysis of the two-class income distribution in the USA in 1983-2018. {\it Phil. Trans. R. Soc.} A 380, 20210162. (doi:10.1098/ rsta.2021.0162) 23.  arXiv:2110.03140v1 [physics.soc-ph] 7 Oct 2021
}

 \end{enumerate}

\end{document}